\documentstyle[aps,multicol,tabularx,rotate,epsf]{revtex}
\begin{document}
\title{Dissipation in Dynamics of a Moving Contact Line}
\author{Ramin Golestanian$^{1,2,3}$ and Elie Rapha\"el$^{1}$}
\address{$^{1}$ Laboratoire de Physique de la Matiere Condensee,
College de France, URA No. 792 du CNRS, 11 place Marcelin-Berthelot,
75231 Paris Cedex 05, France\\
$^{2}$Institute for Advanced Studies in Basic Sciences, Zanjan
45195-159, Iran \\
$^{3}$Institute for Studies in Theoretical Physics and
Mathematics, P.O. Box 19395-5531, Tehran, Iran}
\date{\today}
\maketitle
\begin{abstract}
The dynamics of the deformations of a moving contact line is
studied assuming two different dissipation mechanisms. It is shown
that the characteristic relaxation time for a deformation of
wavelength $2\pi/|k|$ of a contact line moving with velocity $v$
is given as $\tau^{-1}(k)=c(v) |k|$. The velocity dependence of
$c(v)$ is shown to drastically depend on the dissipation
mechanism: we find $c(v)=c(v=0)-2 v$ for the case when the
dynamics is governed by microscopic jumps of single molecules at
the tip (Blake mechanism), and $c(v)\simeq c(v=0)-4 v$ when
viscous hydrodynamic losses inside the moving liquid wedge
dominate (de Gennes mechanism). We thus suggest that the debated
dominant dissipation mechanism can be experimentally determined
using relaxation measurements similar to the Ondarcuhu-Veyssie
experiment [T. Ondarcuhu and M. Veyssie, Nature {\bf 352}, 418
(1991)].
\end{abstract}
\pacs{68.10, 68.45, 05.40}
\begin{multicols}{2}
\section{Introduction}  \label{sIntro}

Spreading of a liquid on a solid surface usually involves a rather
complex dynamical behavior, which is determined by a subtle
competition between the mutual interfacial energetics of the
coexisting phases (the solid, the liquid, and the corresponding
equilibrium vapor), dissipation processes and geometrical or
chemical irregularities of the solid surface \cite{dG1}.
Interestingly, this dynamics can be effectively studied in terms
of the dynamics of the {\it contact line}, which is the common
borderline between the three phases, by correctly taking into
account the physical processes in the vicinity of it.

One of the key issues about this dynamics which has remained a
subject of controversy is dissipation. There are two rival
theories in the literature, each depicting a different physical
picture for the dominant dissipation mechanism in the dynamics of
partial wetting \cite{BWdG}. The first approach, which is based on
the idea of Yarnold and Mason \cite{YM} and was later developed
into a quantitative theory by Blake and coworkers \cite{Blake},
emphasizes the role of microscopic jumps of single molecules (from
the liquid into the vapor) in the immediate vicinity of the
contact line. The other approach, which was developed by de Gennes
and coworkers \cite{dG1,hydro}, asserts that for small values of
contact angle the dissipation is dominated by viscous hydrodynamic
losses inside the moving liquid wedge.

For a partially wetting fluid on sufficiently smooth substrates, a
contact line at equilibrium has a well defined contact angle
$\theta_e$ that is determined by the solid-vapor $\gamma_{SV}$ and
the solid-liquid $\gamma_{SL}$ interfacial energies, and the
liquid surface tension $\gamma$ through Young's relation:
$\gamma_{SV}-\gamma_{SL}=\gamma \cos \theta_e$. For a moving
contact line, however, the value of the so-called dynamic contact
angle $\theta_d$ changes as a function of velocity: $\theta_d >
\theta_e$ for an advancing contact line and $\theta_d < \theta_e$
for a receding one. Since the discrepancy between the two
dissipation mechanisms appears for small contact angles
\cite{BWdG}, one can expect that receding contact lines are in
fact very good candidates for experimental determination of the
dominant mechanism in this regime.

A classic such example corresponds to wetting of a plate that is
vertically withdrawn from a liquid at a constant velocity $-v$,
which was first studied by Landau and Levich for complete wetting
\cite{LL}. In the case of partial wetting that was studied by de
Gennes \cite{hydro,Voinov}, a steady state is achieved in which
the liquid will partially wet the plate with a nonvanishing
dynamic contact angle $\theta_d(v)$ only for pull-out velocities
less than a certain critical value $v_c$. The dynamic contact
angle decreases with increasing $v$, until at the critical
velocity the system undergoes a dynamical phase transition in
which a macroscopic Landau--Levich liquid film, formally
corresponding to a vanishing $\theta_d$, will remain on the plate.

Since the onset of leaving a film occurs at small values of
contact angle, one can imagine that the two different dissipation
mechanisms would lead to conflicting predictions about the
transition. In particular, in Blake's picture the ``order
parameter'' for the transition $\theta_d$ would vanish
continuously as $v$ approaches $v_c$, which makes it look like a
second order phase transition. On the contrary, de Gennes predicts
a jump in the order parameter from $\theta_e/\sqrt{3}$ to zero at
the transition, which is the signature of a first order phase
transition \cite{BWdG,hydro}. This drastic difference in the
predictions of the two theories can provide a reliable venue for
testing them. However, such experiments have so far proven to be
inconclusive due to the usual difficulties of tuning into a
critical point in the presence of disorder \cite{Quere}.

Another notable feature of contact lines is their anomalous
elasticity as noticed by Joanny and de Gennes \cite{JdG1}. For
length scales below the capillary length (which is of the order of
3 mm for water at room temperature), a contact line deformation of
wavevector $k$, denoted as $h(k)$ in Fourier space, will distort
the surface of the liquid over a distance $|k|^{-1}$. Assuming
that the surface deforms instantaneously in response to the
contact line, the elastic energy cost for the deformation can be
calculated from the surface tension energy stored in the distorted
area, and is thus proportional to $|k|$, namely \cite{JdG1}
\begin{equation}
E_{\rm cl}={\gamma \theta_e^2 \over 2} \int {d k \over 2 \pi} |k|
|h(k)|^2.\label{E-cl}
\end{equation}

The anomalous elasticity leads to interesting equilibrium
dynamics, corresponding to when the contact line is perturbed from
its static position, as studied by de Gennes \cite{dG2}. Balancing
${d E_{\rm cl} \over d t}$ and the dissipation, which he assumes
for small contact angles is dominated by the hydrodynamic
dissipation in the liquid nearby the contact line, he finds that
each deformation mode relaxes to equilibrium with a characteristic
inverse decay time $\tau^{-1}(k)=c_0 |k|$, in which $c_0={\gamma
\theta_e^3/(3 \eta \ell)}$ where $\eta$ is the viscosity of the
liquid and $\ell$ is a logarithmic factor of order unity
\cite{dG2}. The relaxation is thus characterized by a linear
dispersion relation, which implies that a deformation in the
contact line will decay and propagate at a constant velocity
$c_0$, as opposed to systems with normal line tension elasticity,
where the decay and the propagation are governed by diffusion.
This behavior has been observed, and the linear dispersion
relation has been precisely tested, in a very interesting
experiment by Ondarcuhu and Veyssie \cite{exp1}.

Here we study the dynamics of the deformations of a moving contact
line for the two different dissipation mechanisms. In particular,
we focus on the sizeable regime where the contact line is moving,
i.e. it is away from the depinning transition \cite{depinning},
but still not too close to the onset of leaving a film. We show
that in this regime the characteristic relaxation time for a
$k$-mode deformation is given as $\tau^{-1}(k)=c(v) |k|$. The
velocity dependence of $c(v)$ is shown to drastically depend on
the dissipation mechanism: we find $c(v)=c(v=0)-2 v$ in Blake's
scheme, whereas in de Gennes' picture $c(v)$ might be rather well
represented by $c(v)\simeq c(v=0)-4 v$. We thus suggest that
monitoring the deformation dynamics in this regime, along the
lines of Ondarcuhu-Veyssie experiment \cite{exp1}, can provide a
more practical probe for experimental determination of the debated
dominant dissipation mechanism.

The rest of the paper is organized as follows. In Sec.
\ref{sDiss}, we discuss the two different dissipation mechanisms
and derive expressions for the corresponding energy dissipation
rates. These expressions are then used in Sec. \ref{sForce} to
derive the force balance, and thus the governing dynamical
equation. The velocity dependence of the dynamic contact angle and
a characteristic velocity are studied in Secs. \ref{sC-V} and
\ref{sChV} correspondingly. While Sec. \ref{sRough} discusses the
effects of surface disorder, we conclude with some discussions in
Sec. \ref{sDis}.

\section{Dissipation}   \label{sDiss}

\begin{figure}
\epsfxsize 7.0cm {\epsffile{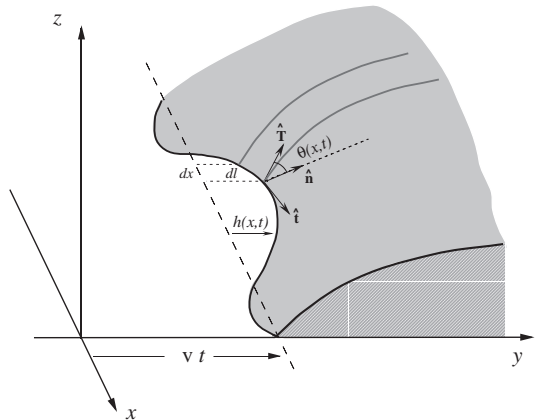}}
\caption{The schematics of the system.} \label{fig:schematics}
\end{figure}

Let us assume that the contact line is directed on average along
the $x$ axis, and is moving in the $y$ direction with an average
velocity $v$, which we assume to be positive corresponding to
receding contact lines, as in Fig.~\ref{fig:schematics}. We can
describe the position of the contact line along the $y$ axis for
any given $x$ and $t$ with the function $y(x,t)=v t+h(x,t)$. We
further assume that the deformation $h(x,t)$ is only a relatively
small perturbation. We can now try to evaluate the overall
dissipation for the deforming contact line within the two
different scenarios.

\subsection{Blake approach}  \label{subB1}

The physical process that is involved in causing dissipation in
Blake's picture, i.e. molecular jumps near the contact line, is
{\em local} in nature \cite{Blake}. Therefore, in any small
neighborhood the amount of dissipation is completely determined by
the local value of the contact line velocity, while all the
molecular details of the dissipation is encoded in an effective
friction coefficient $\mu^{-1}$. The overall dissipation can then
be written as
\begin{equation}
P_{l}={1 \over 2 \mu} \int d x \left[v+\partial_t h(x,t)\right]^2.
\label{P-Bl}
\end{equation}
In the limit of relatively small contact angles, which is relevant
for our receding contact lines, the inverse friction coefficient
(or mobility) can be calculated as \cite{BWdG}
\begin{equation}
\mu=\frac{k \lambda^3}{k_B T} \exp\left(-{W \over k_B
T}\right)\label{Mu},
\end{equation}
in which $W$ is an activation energy for molecular hopping,
$\lambda$ is the distance between hopping sites, $k$ is a
characteristic ``attempt'' rate, and $k_B T$ is the thermal
energy.

\subsection{de Gennes approach}  \label{subPGG1}

We now focus on the contribution of dissipation that comes from
the viscous losses in the hydrodynamic flows inside the liquid
wedge \cite{dG1,hydro}. For a slightly deformed contact line, we
assume that the dissipation can be approximated by the sum of
contributions from wedge-shaped slices with local contact angles
$\theta(x,t)$, as shown in Fig.~\ref{fig:schematics}. This is a
reasonable approximation because most of the dissipation is taking
place in the singular flows near the tip of the wedge
\cite{dG1,hydro,dG2}. Using the result for the dissipation in a
perfect wedge, which is based on the lubrication approximation
\cite{dG1,Huh}, we can calculate the total dissipation as
\cite{dG2}
\begin{equation}
P_{h}={\eta \over 2}\int d x  \left({3 \ell \over
\theta(x,t)}\right) \left[v+\partial_t h(x,t)\right]^2,
\label{P-PGG}
\end{equation}
in which $\ell=\ln(d_{max}/d_{min})$ with $d_{max}$ given by the
size of the liquid drop and $d_{min}$ being a microscopic length
scale. The inverse dependence on $\theta$ suggests that for
sufficiently small contact angles the hydrodynamic loss is to be
dominant \cite{BWdG}.

\section{Force balance and dynamical equation}  \label{sForce}

To find the governing dynamical equation in the long time limit,
we should balance the total friction force obtained as $\delta
(P_l+P_h)/\delta \partial_t h(x,t)$ with the interfacial force
$\gamma \cos \theta(x,t)-(\gamma_{SV}-\gamma_{SL})=\gamma [\cos
\theta(x,t)-\cos \theta_e]$ at each point along the contact line.
Note that in this section we are taking both dissipation
mechanisms into account. In the limit of small contact angles, we
find
\begin{equation}
\left[{1 \over \mu}+{3 \eta \ell \over \theta(x,t)}\right] \times
\left[v+\partial_t h(x,t)\right]={\gamma \over 2}
\left[\theta_e^2-\theta(x,t)^2\right]. \label{D-E}
\end{equation}
To proceed from here, we need to relate the contact angle
$\theta(x,t)$ to the contact line profile $h(x,t)$, which can be
done through solving for the surface profile of the liquid drop.
One can show that the surface profile $z(x,y)$ near the contact
line can be found as a solution of the Laplace equation
$(\partial^2_x+\partial^2_y)z(x,y)=0$, so as to minimize the
surface area. The solution that satisfies the boundary condition
$z(x,h(x,t))=0$ reads \cite{JdG1}
\begin{equation}
z(x,y)=\theta_d \left[y-\int {d k \over 2 \pi} h(k,t) e^{i k x-|k|
y}\right],\label{profile}
\end{equation}
from which we obtain
\begin{equation}
\theta(x,t)\equiv \left.{\partial z(x,y) \over \partial y
}\right|_{y=h(x,t)}=\theta_d \left[1+\int {d k \over 2 \pi} |k|
h(k,t) e^{i k x}\right],\label{theta(x,t)}
\end{equation}
to the leading order \cite{JdG1}.

To the zeroth order, Eq.(\ref{D-E}) gives the relation between the
average dynamic contact angle and the velocity as
\begin{equation}
v=\left(\gamma \over 6 \eta \ell \right) {\theta_d
(\theta_e^2-\theta_d^2) \over 1+{\theta_d/(3 \eta \mu \ell)}}.
\label{v-theta}
\end{equation}
This relation will be used below to study the onset of the
transition of the moving liquid drop to a Landau-Levich film.

The dynamical equation (Eq.(\ref{D-E})), which governs the
dynamics of the deformation field, can now be written in the
linear approximation as
\begin{equation}
\partial_t h(k,t)=-c(v) |k| h(k,t),  \label{relax}
\end{equation}
in Fourier space, where
\begin{equation}
c(v)={\mu \gamma \theta_d^3(v)- 3 \eta \mu \ell v \over 3 \eta \mu
\ell+\theta_d(v)}, \label{c(v)1}
\end{equation}
with $\theta_d(v)$ to be found by inverting Eq.(\ref{v-theta}).
The corresponding form of the dynamical equation in real space can
be found by Fourier transformation as
\begin{equation}
\partial_t h(x,t)=-c(v) \int {d x' \over \pi} {h(x',t) \over (x-x')^2},\label{relax2}
\end{equation}
which reflects the non-locality of the dynamics.

Relaxation of the contact line's shape while it is moving thus
takes place with the same dispersion relation $\tau^{-1}(k) \sim
|k|$ as a contact line at rest, although the corresponding {\it
characteristic velocity} $c(v)$ shows a strong dependence on the
contact line velocity $v$.

\section{Contact angle--velocity relation}  \label{sC-V}

There can be two types of experiments on a moving contact line
depending on how we prepare it. We can fix a value for the contact
angle that is different from $\theta_e$, and let it move with an
adjusted velocity when it reaches a steady state. This can be
achieved, for example, by adding or removing some volume of liquid
through a syringe that is inserted in a liquid drop at
equilibrium. On the contrary, we can fix the velocity and let the
contact angle adjust itself in a steady state. This will be the
case, for example, when a plate is withdrawn vertically from a
liquid at a constant velocity.

Depending on which ``ensemble'' we are using, we will have a fixed
value for $v$ or $\theta_d$, and we should then use
Eq.(\ref{v-theta}) (that relates the velocity and the dynamic
contact angle) to determine the conjugate parameter. The term
``ensemble'' is in fact quite appropriate to use here because the
two (mechanically) conjugate quantities are, in fact, velocity and
force--which is determined solely by the contact angle. What we
have is then either a ``constant velocity'' or a ``constant
force'' experiment. It is interesting to note that in
non-equilibrium systems, in general, different ensembles may not
necessarily lead to the same result \cite{ensemble}.

In this work, we are mostly interested in constant velocity
experiments, and thus we will treat $v$ as a fixed and given
parameter below unless otherwise specified. We will examine
Eq.(\ref{v-theta}) in the limiting cases corresponding to the two
different dissipation mechanisms and compare their predictions.

\subsection{Blake approach}  \label{subB2}

\begin{figure}
\centerline{\epsfxsize 7cm {\epsffile{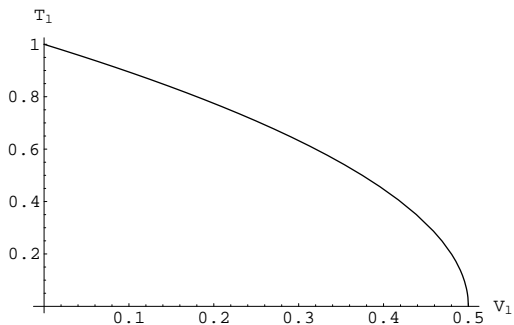}}}
\vskip.3truecm \caption{The reduced order parameter
$T_l=(\theta_d/\theta_e)_l$ as a function of the dimensionless
velocity $V_l=v/c_{l0}$ for Blake mechanism (Eq.(\ref{cav-l})).
The dynamical phase transition at $V_{lc}=1/2$ is predicted to be
of second order in this picture.} \label{fig:theta-l}
\end{figure}

The behavior in this regime can be extracted from
Eq.(\ref{v-theta}) by taking the limit $\mu \eta \ll \theta_d$.
Inverting the resulting equation yields
\begin{equation}
\left.\theta_d(v) \over \theta_e\right|_l=\sqrt{1-{2 v \over
c_{l0}}}, \label{cav-l}
\end{equation}
in which $c_{l0}=\mu \gamma \theta_e^2$. Note that this holds only
for $v< c_{l0}/2$, while $\theta_d=0$ identically for
$v>c_{l0}/2$. This function is plotted in Fig.~\ref{fig:theta-l}.

As can be readily seen from Fig.~\ref{fig:theta-l}, increasing $v$
would lead to decreasing values of $\theta_d$ until at a critical
velocity $v_{lc}=c_{l0}/2$ it finally vanishes continuously. A
vanishing contact angle presumably corresponds to formation of a
liquid film; a so-called Landau-Levich film. The value of the
dynamic contact angle $\theta_d$ serves as the order parameter for
this dynamical phase transition, while $v$ is the tuning
parameter. The continuous vanishing of the order parameter makes
the phase transition classified as second order. As in the general
theory of critical phenomena, a mean-field exponent $\beta=1/2$ is
characterizing the vanishing of the order parameter in terms of
the tuning parameter.

\subsection{de Gennes approach}  \label{subPGG2}

In the opposite limit of $\mu \eta \gg \theta_d$, only the
hydrodynamic contribution survives, and Eq.(\ref{v-theta})
leads to
\begin{eqnarray}
\left.\theta_d(v) \over \theta_e\right|_h&=&{1 \over \sqrt{3}}
\left[\left(-\nu-i \sqrt{1-\nu^2}\right)^{1/3}\right. \nonumber \\
&&\left.+\left(-\nu+i \sqrt{1-\nu^2}\right)^{1/3}\right],
\label{cav-h}
\end{eqnarray}
in which $\nu=3 \sqrt{3} v/c_{h0}$ and $c_{h0}=\gamma
\theta_e^3/(3 \eta \ell)$.\footnote{Note that the expression in
Eq.(\ref{cav-h}) is real, and the $i$ is retained only to keep the
appearance of the formula simpler.}

\begin{figure}
\centerline{\epsfxsize 7cm {\epsffile{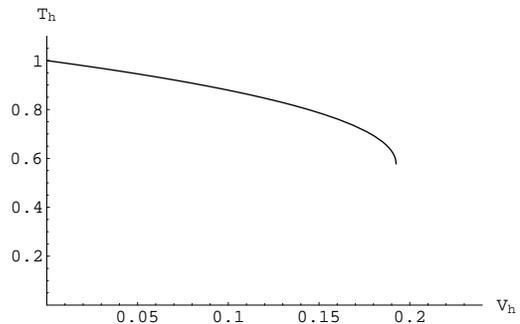}}}
\vskip.3truecm \caption{The reduced order parameter
$T_h=(\theta_d/\theta_e)_h$ as a function of the dimensionless
velocity $V_h=v/c_{h0}$ for de Gennes mechanism
(Eq.(\ref{cav-h})). The dynamical phase transition at
$V_{hc}=1/(3\sqrt{3})\simeq 0.192$ is predicted to be of first
order in this picture.} \label{fig:theta-h}
\end{figure}

The above formula, which holds only for $v<c_{h0}/(3\sqrt{3})$ has
two branches and only the one that recovers $\theta_d(0)=\theta_e$
is acceptable as plotted in Fig.\ref{fig:theta-h}. While at
$v=c_{h0}/(3\sqrt{3})$ we find $\theta_d=\theta_e/\sqrt{3}$, we
expect to have $\theta_d=0$ for higher velocities
$v>c_{h0}/(3\sqrt{3})$. Therefore, the order parameter $\theta_d$
experiences a finite jump at the transition velocity
$v_{hc}=c_{h0}/(3\sqrt{3})$, that is the hallmark of a first order
phase transition.

\section{Characteristic velocity}   \label{sChV}

Using Eq.(\ref{c(v)1}), and $\theta_d(v)$ that we have found in
the previous section for the two different cases, we can extract
the $v$-dependence of the characteristic velocity.

\subsection{Blake approach}  \label{subB3}

\begin{figure}
\centerline{\epsfxsize 7cm {\epsffile{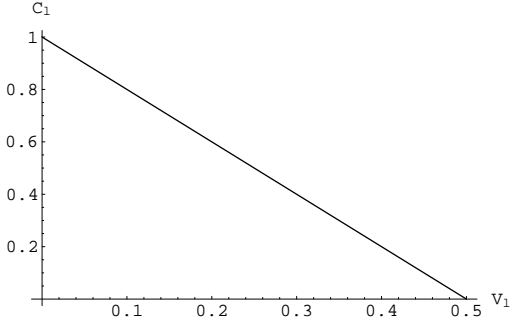}}} \vskip.3truecm
\caption{The reduced characteristic velocity $C_l=c_l/c_{l0}$ as a
function of the dimensionless velocity $V_l=v/c_{l0}$ for Blake
mechanism (Eq.(\ref{cv-l})). The slope of the curve is $-2$ all
the way to the transition point where the characteristic velocity
vanishes.} \label{fig:c-l}
\end{figure}

We can simplify Eq.(\ref{c(v)1}) by taking the limit $\mu \eta \ll
\theta_d$, as
\begin{equation}
c_l(\theta)=c_{l0} {\theta_d^2 \over \theta_e^2}.\label{ctheta-l}
\end{equation}
Inserting the form of $\theta_d(v)$ from Eq.(\ref{cav-l}) yields
\begin{equation}
c_l(v)=c_{l0}-2 v.\label{cv-l}
\end{equation}
Note that in this approach $c(v)$ is strictly linear in $v$ all
the way, and it vanishes at the transition point, as plotted in
Fig.~\ref{fig:c-l}.

\subsection{de Gennes approach}  \label{subPGG3}

\begin{figure}
\centerline{\epsfxsize 7cm {\epsffile{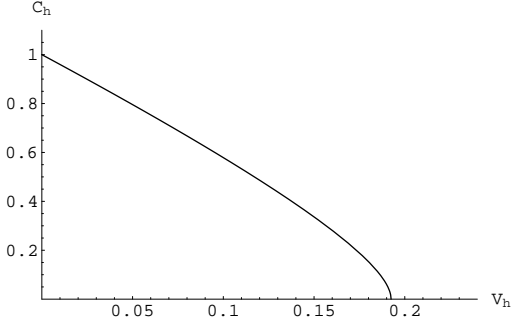}}} \vskip .3truecm
\caption{The reduced characteristic velocity $C_h=c_h/c_{h0}$ as a
function of the dimensionless velocity $V_h=v/c_{h0}$ for de
Gennes mechanism (Eq.(\ref{cv-h})). The slope of the curve is
nearly $-4$ until a square-root singularity sets in near the
transition point where the characteristic velocity vanishes.}
\label{fig:c-h}
\end{figure}

In the opposite limit of $\mu \eta \gg \theta_d$, Eq.(\ref{c(v)1})
will be simplified as
\begin{equation}
c_h(\theta)={c_{h0} \over 2} \left(3 {\theta_d^3 \over
\theta_e^3}-{\theta_d \over \theta_e} \right).\label{ctheta-h}
\end{equation}
Putting in $\theta_d(v)$ from Eq.(\ref{cav-h}) leads to
\begin{eqnarray}
c_h(v)&=&{c_{h0}\over \sqrt{3}} \left[\left(-\nu-i
\sqrt{1-\nu^2}\right)^{1/3}\right.\nonumber \\
&&\left.+ \left(-\nu+i
\sqrt{1-\nu^2}\right)^{1/3}-\nu\right].\label{cv-h}
\end{eqnarray}
One can again check from this equation that $c(v)$ vanishes at the
transition. The above equation is plotted in Fig.~\ref{fig:c-h}.

The characteristic velocity can be well approximated by the linear
expression
\begin{equation}
c_h(v)\simeq c_{h0}-4 v,\label{cv0-h}
\end{equation}
for a wide range of $v$, except very near $c_{h0}/(3\sqrt{3})$
where it experiences a square-root singular behavior as
\begin{eqnarray}
c_h(v)&\simeq& c_{h0} \left[\sqrt{2}\left({1 \over 3 \sqrt{3} }-{v
\over c_{h0}}\right)^{1/2}+3\left({1 \over 3 \sqrt{3} }-{v \over
c_{h0}}\right)\right. \nonumber \\
&&\left.+O\left(\left({1 \over 3 \sqrt{3} }-{v \over
c_{h0}}\right)^{3/2}\right)\right].\label{cv1-h}
\end{eqnarray}
It is interesting to note that although both approaches predict a
sizeable linear regime for $c(v)$, as manifest in Eqs.(\ref{cv-l})
and (\ref{cv0-h}), the corresponding slopes are predicted
differently.

\section{Surface Disorder}  \label{sRough}

In most practical cases, the dynamics of a contact line is
affected by the defects and heterogeneities in the substrate, in
addition to dissipation and elasticity that we have considered so
far. If the interfacial energies $\gamma_{SV}$ and $\gamma_{SL}$
are space dependent with the corresponding averages being ${\bar
\gamma_{SV}}$ and ${\bar \gamma_{SL}}$, a displacement $\delta
y(x,t)$ of the contact line is going to lead to a change in energy
as
\begin{equation}
\delta E_{d}=\int d x g(x,v t+h(x,t)) \delta y(x,t),\label{dE}
\end{equation}
where
\begin{equation}
g(x,y)=\gamma_{SV}(x,y)-\gamma_{SL}(x,y)-({\bar \gamma_{SV}}-{\bar
\gamma_{SL}}).\label{g-def}
\end{equation}
Incorporating this contribution in the force balance leads to an
extra force term $g(x,v t)$ on the right hand side of
Eq.(\ref{D-E}), and thus a noise term on the right hand side of
Eq.(\ref{relax2}) of the form
\begin{equation}
\eta(x,t)=\left({\mu \theta_d \over \theta_d+3 \eta \mu
\ell}\right) g(x,vt),\label{eta-def}
\end{equation}
to the leading order. Note that this is a good approximation
provided we are well away from the depinning transition, and the
contact line is moving fast enough \cite{dG1,JdG1,depinning,RJ1}.

Assuming that the surface disorder has short range correlations
with a Gaussian distribution described by
\begin{eqnarray}
\langle g(x,y) \rangle&=&0, \nonumber \\
\langle g(x,y) g(x',y')\rangle&=&g^2 a^2\delta(x-x')
\delta(y-y'),\label{g-moments}
\end{eqnarray}
we can deduce the distribution of the noise as
\begin{eqnarray}
\langle \eta(x,t) \rangle&=&0, \nonumber \\
\langle \eta(x,t) \eta(x',t')\rangle &=& 2 D(v) \delta(x-x')
\delta(t-t'),\label{eta-moments}
\end{eqnarray}
where
\begin{equation}
D(v)={g^2 a^2 \over 2 v}\left({\mu \theta_d(v) \over \theta_d(v)+3
\eta \mu \ell}\right)^2.\label{D-def}
\end{equation}
In the presence of the noise, the contact line undergoes dynamical
fluctuations. These fluctuations can best be characterized by the
width of the contact line, which is defined as
\begin{equation}
W^2(L,t)\equiv {1 \over L} \int d x \langle h(x,t)^2
\rangle.\label{W-def}
\end{equation}
Using Eq.(\ref{relax}) with the noise term, we can calculate the
width of the contact line as
\begin{eqnarray}
W^2(L,t)&=&{D(v) \over \pi c(v)}\times \int_{\pi/L}^{\pi/a} {d k
\over k} \left[1-e^{-2 c(v) |k|t }\right] \nonumber \\
&=&{D(v) \over \pi c(v)} \times \left\{\begin{array}{ll}
\ln\left[c(v) t/a\right], &\; {a \over c(v)} \ll t \ll {L \over
c(v)},  \\
\ln\left(L/a\right), &\; t \gg {L \over c(v)},
\end{array} \right. \label{W1}
\end{eqnarray}

Similarly, we can study the fluctuations in the order parameter
field $\delta \theta(x,t)=\theta(x,t)-\theta_d$. Using
Eqs.(\ref{theta(x,t)}) and (\ref{relax}), we find
\begin{eqnarray}
\langle \delta \theta(x,t)^2 \rangle&=&{D(v) \theta_d^2(v)\over
\pi c(v)} \times \int_{\pi/L}^{\pi/a} d k \; k \left[1-e^{-2 c(v) |k|t }\right] \nonumber \\
&=&{\pi D(v) \theta_d^2(v)\over 2 c(v) a^2} \left(1-{a^2 \over 2
\pi^2 c^2(v) t^2}\right), \label{theta0}
\end{eqnarray}
for $t \gg a/c(v)$.

The magnitude of the fluctuations of the contact line width
\begin{equation}
\Delta(v)={D(v) \over \pi c(v)},\label{Delta-def}
\end{equation}
and, correspondingly, that of the order parameter
\begin{equation}
\sigma(v)={\pi D(v) \theta_d^2(v)\over 2 c(v)
a^2},\label{Sigma-def}
\end{equation}
are thus both velocity dependent. Again, we expect this dependence
to be different for the two cases.

\subsection{Blake approach}  \label{subB4}

\begin{figure}
\centerline{\epsfxsize 7cm {\epsffile{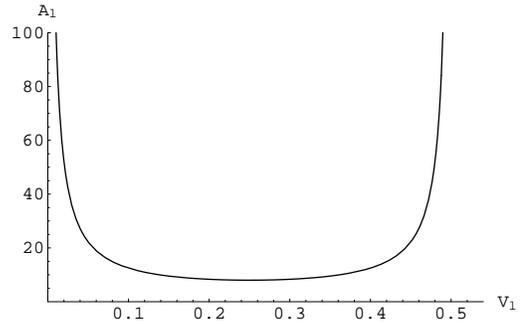}}} \vskip
.3truecm \caption{The reduced width $A_l=\Delta_l/\left({\mu^2 g^2
a^2\over 2 \pi c_{l0}^2}\right)$ as a function of the
dimensionless velocity $V_l=v/c_{l0}$ for Blake mechanism
(Eq.(\ref{Delta-l})). It is symmetric with respect to $V_l=1/4$.}
\label{fig:delta-l}
\end{figure}

\begin{figure}
\centerline{\epsfxsize 7cm {\epsffile{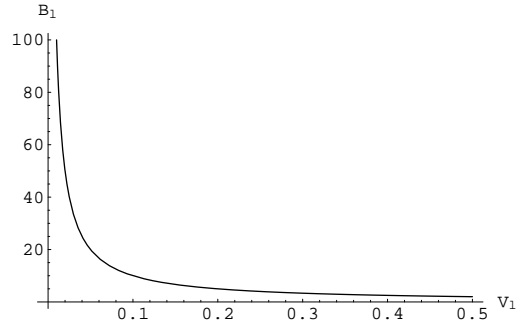}}} \vskip
.3truecm \caption{The reduced order parameter fluctuations
$B_l=\sigma_l/\left({\pi \mu^2 g^2 \theta_e^2 \over 4 c_{l0}^2
}\right)$ as a function of the dimensionless velocity
$V_l=v/c_{l0}$ for Blake mechanism (Eq.(\ref{Sigma-l})). It is a
monotonically decreasing function of velocity. Note the unusual
feature that the order parameter fluctuations {\em remain finite}
at the transition point, unlike traditional second order phase
transitions.} \label{fig:sigma-l}
\end{figure}

In this case we have $\mu \eta \ll \theta_d$, which together with
Eqs.(\ref{cav-l}), (\ref{cv-l}), (\ref{D-def}), (\ref{Delta-def}),
and (\ref{Sigma-def}) yield
\begin{equation}
\Delta_{l}(v)=\left({\mu^2 g^2 a^2\over 2 \pi c_{l0}^2}\right)
\times {1 \over (v/c_{l0})(1-2 v/c_{l0})},\label{Delta-l}
\end{equation}
and
\begin{equation}
\sigma_{l}(v)=\left({\pi \mu^2 g^2 \theta_e^2 \over 4 c_{l0}^2
}\right)\times {1 \over (v/c_{l0})}.\label{Sigma-l}
\end{equation}
The above equations are plotted in Figs.~\ref{fig:delta-l} and
\ref{fig:sigma-l}. Fig.~\ref{fig:delta-l} shows that the width of
the contact line is a symmetric function of velocity in this
picture, while Fig.~\ref{fig:sigma-l} denotes that the order
parameter fluctuations decrease monotonically with velocity. Note
also that these fluctuations remain finite at the transition
point, which is not typical of second order phase transitions.

\subsection{de Gennes approach}  \label{subPGG4}

\begin{figure}
\centerline{\epsfxsize 7cm {\epsffile{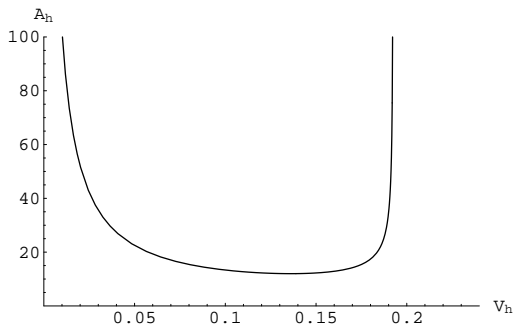}}} \vskip
.3truecm \caption{The reduced width $A_h=\Delta_h/\left({g^2 a^2
\theta_e^2\over 18 \pi \eta^2 \ell^2 c_{h0}^2 }\right)$ as a
function of the dimensionless velocity $V_h=v/c_{h0}$ for de
Gennes mechanism (Eq.(\ref{Delta-h})). Note the asymmetry of the
plot in this case.} \label{fig:delta-h}
\end{figure}

\begin{figure}
\centerline{\epsfxsize 7cm {\epsffile{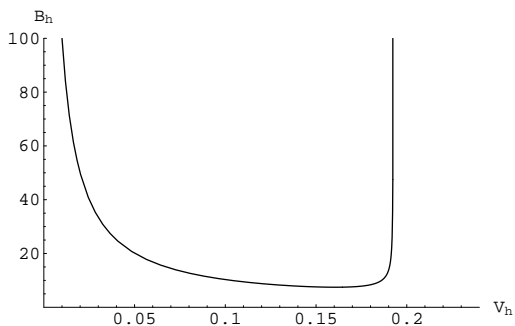}}} \vskip
.3truecm \caption{The reduced order parameter fluctuations
$B_h=\sigma_h/\left({\pi g^2 \theta_e^4\over 36 \eta^2 \ell^2
c_{h0}^2 }\right)$ as a function of the dimensionless velocity
$V_h=v/c_{h0}$ for de Gennes mechanism (Eq.(\ref{Sigma-h})). In
this case, it is {\em not} a monotonically decreasing function of
velocity. Note the unusual feature that the order parameter
fluctuations {\em diverge} at the transition point, unlike
traditional first order phase transitions.} \label{fig:sigma-h}
\end{figure}

Taking the opposite limit $\mu \eta \gg \theta_d$ in
Eq.(\ref{D-def}), together with Eqs.(\ref{cav-h}), (\ref{cv-h}),
(\ref{Delta-def}), and (\ref{Sigma-def}), we obtain
\end{multicols}
\begin{equation}
\Delta_{h}(v)=\left({g^2 a^2 \theta_e^2\over 6 \pi \eta^2 \ell^2
c_{h0}^2 }\right)\times \left\{\frac{\left[\left(-\nu-i
\sqrt{1-\nu^2}\right)^{1/3}+\left(-\nu+i
\sqrt{1-\nu^2}\right)^{1/3}\right]^2}{\nu \left[\left(-\nu-i
\sqrt{1-\nu^2}\right)^{1/3}+\left(-\nu+i
\sqrt{1-\nu^2}\right)^{1/3}-\nu\right]}\right\},\label{Delta-h}
\end{equation}
and
\begin{equation}
\sigma_{h}(v)=\left({\pi g^2 \theta_e^4\over 36 \eta^2 \ell^2
c_{h0}^2 }\right)\times \left\{\frac{\left[\left(-\nu-i
\sqrt{1-\nu^2}\right)^{1/3}+\left(-\nu+i
\sqrt{1-\nu^2}\right)^{1/3}\right]^4}{\nu \left[\left(-\nu-i
\sqrt{1-\nu^2}\right)^{1/3}+\left(-\nu+i
\sqrt{1-\nu^2}\right)^{1/3}-\nu\right]}\right\}.\label{Sigma-h}
\end{equation}
\begin{multicols}{2}\noindent
The above equations are plotted in Figs.~\ref{fig:delta-h} and
\ref{fig:sigma-h}. Fig.~\ref{fig:delta-h} shows that the width of
the contact line is not a symmetric function of velocity in this
case. Moreover, the order parameter fluctuations do not decrease
monotonically with velocity as shown in Fig.~\ref{fig:sigma-h}.
Unlike in the previous case, these fluctuations diverge at the
transition point, which is again not typical of first order phase
transitions.

\section{Discussion}  \label{sDis}

Because of their anomalous elasticity, contact lines relax to
their equilibrium from an initially distorted configuration with a
characteristic inverse decay time $\tau^{-1}(k)=c(v) |k|$ for each
$k$-mode. The $v$-dependence of the characteristic velocity $c(v)$
is shown to crucially depend on the dissipation mechanism, and it
can thus be used as an experimental probe for the dominant
dissipation mechanism.

A typical experiment for such investigations is direct monitoring
of the contact line shape during the relaxation process. If the
initial distortion of the contact line can be made periodic in a
controlled way, like in the experiment of Ondarcuhu and Veyssie
\cite{exp1}, one can directly map out $c(v)$ and hence determine
the dissipation mechanism from its $v$-dependence.

Another possibility is to have relaxation from random initial
distortions, which will be the case when we pull out a naturally
rough plate from the liquid. Monitoring the dynamics of the
contact line in this case will provide statistical information
about the relaxation process, from which one can hope to deduce
the relevant features discussed in Sec. \ref{sRough}.

We finally note that this linear theory is not sufficient for a
complete understanding of the Landau-Levich phase transition, and
it breaks down upon approaching the transition point. This
breakdown is particularly manifest in the divergence that we
encountered in the width of the contact line at the transition
point. To be able to have a complete description, one should keep
the relevant nonlinear terms that can be calculated by extending
the method of this paper, and resort to perturbative
renormalization group approaches for the resulting nonlinear
stochastic equation. We have performed these studies, and the
corresponding results will appear elsewhere \cite{RE}.

In conclusion, we have studied the relaxation dynamics of the
contact lines, and suggested that monitoring this dynamics can
provide an experimental probe for the debated dominant dissipation
mechanism.

\acknowledgments

We are grateful to J. Bico, P.G. de Gennes, and D. Qu\'er\'e for
invaluable discussions and comments. One of us (R.G.) would like
to thank the group of Prof. de Gennes at College de France for
their hospitality and support during his visit. This research was
supported in part by the National Science Foundation under Grant
No. DMR-98-05833 (R.G.).

\end{multicols}
\end{document}